\documentclass[11pt]{article}
\usepackage{amscd}
\usepackage{amssymb}
\usepackage{amsfonts}
\usepackage{amsmath}
\usepackage{mathrsfs}
\begin{document}

\setcounter{page}{1}
\vfil

\pagestyle{plain}

\begin{center}
{\LARGE {\bf Coadjoint Orbits, Spin and Dequantization}}

\bigskip
\bigskip
\bigskip

{\large D. Mauro}

             Department of Theoretical Physics, 
	     University of Trieste, \\
	     Strada Costiera 11, Miramare-Grignano, 34014 Trieste, Italy,\\
	     and INFN, Trieste, Italy\\
	     e-mail: {\it mauro@ts.infn.it}

\end{center}

\bigskip
\bigskip
\begin{abstract}
In this Letter we propose two path integral approaches to describe the
classical mechanics of spinning particles. We show how these formulations can be derived from the associated quantum ones via a sort of geometrical dequantization procedure proposed in a previous paper. 
\end{abstract}

\section{Introduction}

Feynman's path integral is one of the most fruitful methods to study quantum mechanics.
Nevertheless in Ref. \cite{Feyn} R. P. Feynman himself said that ``{\it path integrals
suffer most grievously from a serious defect. They do not permit a discussion of spin operators}". The reason for this difficulty is that the path integral formulation needs as an ingredient the Lagrangian of the system, which is a classical concept, and nothing like that existed for the spin in the Forties and the Fifties. 
Since then this problem has been overcome. Various ideas \cite{annuno}, \cite{Casalbuoni}, \cite{Berezin}, \cite{nielsen}, \cite{johnson} to formulate quantum path integrals for spinning particles have been put forward. 
These ideas can be divided in two main lines of thought. The first one
goes as follows: since spinning particles are described by Pauli matrices, which are anticommuting operators, the underlying classical mechanics must be formulated via anticommuting 
or Grassmann numbers. Casalbuoni and independently Berezin and Marinov went into 
this direction in \cite{Casalbuoni}-\cite{Berezin} and their path integral for spinning particles involves a functional integration over Grassmann variables. 
Another quantum path integral formulation for particles with spin, described in 
Refs. \cite{nielsen}-\cite{johnson},
involves instead a functional integration over a set of bosonic phase space variables 
whose choice is dictated by the symplectic form associated with the coadjoint orbits
of the SO(3) group \cite{alek}. The weight appearing in these two quantum path integrals is given by two Lagrangians which describe the spin degrees of freedom. By minimizing the action associated to these Lagrangians one gets two ``{\it classical}" descriptions of the spin. This may sound quite strange because most people think that spin is an intrinsically {\it quantum} concept. This is actually wrong. It is known in fact that the concept of spin appears not only in the {\it quantum} unitary representations of SO(3), but also in the canonical realizations of this group \cite{pauri}, which are intrinsically {\it classical} constructions.

In this Letter we will explore a third way to get a ``classical" description of spin. This third way is based on the fact that not only {\it quantum} mechanics \cite{Feyn}, but also {\it classical} mechanics can have a path integral formulation \cite{gozzi}. We will indicate this last one with the acronym CPI, for Classical Path Integral, while the Quantum Path Integral will be indicated with QPI. Recently \cite{deq} a dequantization procedure to pass from the QPI to the CPI has been put forward. This dequantization procedure will be our way of getting a {\it classical} description of spin starting from the quantum one. 

The paper is organized as follows: in {\bf Sec. 2} we will give a brief summary of the geometrical dequantization procedure proposed in \cite{deq} for particles without spin;
in {\bf Sec. 3} we shall review the path integral over Grassmann variables that can be used to describe the quantum motion of a particle with spin. In {\bf Sec. 4} we will build the associated CPI, showing that it can be derived via the dequantization procedure. In {\bf Sec. 5} 
we prove that the same ``dequantization" procedure can be applied also to the path integral over bosonic variables developed in \cite{nielsen}-\cite{johnson}.

\section{Geometrical Dequantization for Particles without Spin}

First of all, let us briefly review the basic steps of the dequantization procedure mentioned above, which starts from a reformulation of classical mechanics based on the functional techniques of Ref. \cite{gozzi}. Also at the classical level we can talk \cite{kvn} of probability amplitude $K(\phi^a;t| \phi^a_{\scriptscriptstyle 0};t_{\scriptscriptstyle 0})$ of finding a particle in the point $\phi^a$ 
of the phase space at time $t$ if it was at $\phi_{\scriptscriptstyle 0}^a$ at time 
$t_{\scriptscriptstyle 0}$. This probability amplitude is given by:
\begin{equation}
K(\phi^a;t| \phi^a_{\scriptscriptstyle 0};t_{\scriptscriptstyle 0})=
\int {\mathscr D}^{\prime\prime} \phi \, \widetilde{\delta}\Bigl[\phi^a-\phi^a_{\textrm{cl}}(t;
\phi_{\scriptscriptstyle 0},t_{\scriptscriptstyle 0})\Bigr]
\label{uno}
\end{equation}
where $\phi^a_{\textrm{cl}}$ is the solution of the classical equations of motion $\dot{\phi}^a=\omega^{ab}\partial_bH$ and the symbol ${\mathscr D}^{\prime\prime}\phi$ indicates that the integration is over paths with fixed end points $\phi_{\scriptscriptstyle 0}$ and $\phi$. The functional Dirac delta in (\ref{uno}) can be rewritten as follows \cite{gozzi}:
\begin{equation}
\widetilde{\delta}\Bigl[\phi^a-\phi^a_{\textrm{cl}}(t;\phi_{\scriptscriptstyle 0},t_{\scriptscriptstyle 0})\Bigr]
=\widetilde{\delta}\Bigl[\dot{\phi}^a-\omega^{ab}\partial_bH\Bigr]\,\textrm{det}
\Bigl(\delta_b^a\partial_t-\omega^{ac}\partial_c\partial_bH\Bigr). \label{due}
\end{equation}
We can then exponentiate the functional Dirac delta of the equations of motion 
via the bosonic variables $\lambda_a$
and the functional determinant via the Grassmann variables $c^a$ and $\bar{c}_a$. Consequently
the probability amplitude (\ref{uno}) can be rewritten as the following path integral:
\begin{equation}
K(\phi^a;t|\phi^a_{\scriptscriptstyle 0};t_{\scriptscriptstyle 0})=
\int {\mathscr D}^{\prime\prime}\phi{\mathscr D}\lambda 
{\mathscr D}c{\mathscr D}\bar{c}\; \textrm{exp}\biggl[i
\int_{t_0}^{t} d\tau\, \widetilde{\cal L}\biggr] \label{buno}
\end{equation}
where $\widetilde{\cal L}$ is the following Lagrangian:
\begin{equation}
\widetilde{\cal L}=\lambda_a\dot{\phi}^a+i\bar{c}_a\dot{c}^a-\lambda_a\omega^{ab}\partial_bH
-i\bar{c}_a\omega^{ad}\partial_d\partial_bHc^b. \label{bdue}
\end{equation}
From (\ref{buno}) and the form of the kinetic terms in the Lagrangian (\ref{bdue}) we can derive that the only graded
commutators different from zero are $[\hat{\phi}^a,\hat{\lambda}_b]=i\delta_b^a$ 
and $[\hat{c}^a,\hat{\bar{c}}_b]=
\delta_b^a$. So the operators $\hat{\phi}$ and $\hat{c}$ commute and they can
be diagonalized simultaneously:
\begin{equation}
\label{alfa}
\left\{
\begin{array}{l}
\hat{\phi}\,|\phi,c\rangle=\phi\,|\phi,c\rangle \medskip \\
\hat{c}\,|\phi,c\rangle=c\,|\phi,c\rangle.
\end{array}
\right.
\end{equation}
Therefore the kernel $K(\phi^a,c^a;t|\phi^a_{\scriptscriptstyle 0},c^a_{\scriptscriptstyle 0};t_{\scriptscriptstyle 0})$ can be written as $\langle \phi,c;t|\phi_{\scriptscriptstyle 0}, c_{\scriptscriptstyle 0};t_{\scriptscriptstyle 0}\rangle$ and it has 
the following expression:
\begin{equation}
\displaystyle \langle \phi,c;t|
\phi_{\scriptscriptstyle 0},c_{\scriptscriptstyle 0};t_{\scriptscriptstyle 0}\rangle
=\int {\mathscr D}^{\prime\prime}\phi{\mathscr D}\lambda{\mathscr D}^{\prime\prime}c{\mathscr D}\bar{c}\;
\textrm{exp}\biggl[i\int_{t_0}^{t} d\tau\,\widetilde{\cal L}\biggr]. \label{gamma}
\end{equation}
This path integral is the functional counterpart of the Koopman-von Neumann operatorial approach to classical mechanics \cite{koopman}. It basically reproduces the kernel of evolution associated with a generalization of the Liouville equation for classical statistical mechanics, see \cite{gozzi} for further details. From a geometrical point of view, the weight appearing in the path integral (\ref{gamma}) is related to the Lie derivative of the Hamiltonian flow \cite{gozzi}-\cite{marsden}. 
At first sight the path integral (\ref{gamma}) seems to be completely different from the  QPI:
\begin{equation}
\displaystyle 
\langle q;t|q_{\scriptscriptstyle 0};t_{\scriptscriptstyle 0}\rangle =\int
{\mathscr D}^{\prime\prime}q {\mathscr D}p \; \textrm{exp}\biggl[\frac{i}{\hbar} 
\int_{t_0}^{t}  d\tau\, L(q,p)\biggr] \label{delta}
\end{equation}
where $L(q,p)=p\dot{q}-H(q,p)$. We will show that it is not so. If we actually introduce, besides the time $t$, two Grassmann partners of $t$ called $\theta,\bar{\theta}$ then we can assemble all the $8n$ variables $(\phi^a,\lambda_a,c^a,\bar{c}_a)$ of the path integral (\ref{gamma}) into the following functions of $t$, $\theta$ and $\bar{\theta}$, which are known in the literature on supersymmetry as superfields:
\begin{equation}
\displaystyle 
\left\{
\begin{array}{l}
Q(t,\theta,\bar{\theta})=q(t)+\theta c^q(t)+\bar{\theta}\bar{c}_p(t)+i\bar{\theta}\theta\lambda_p(t)
\smallskip\\
\displaystyle 
P(t,\theta,\bar{\theta})=p(t)+\theta c^p(t)-\bar{\theta}\bar{c}_q(t)-i\bar{\theta}\theta\lambda_q(t).
\label{fi}
\end{array}
\right.
\end{equation}
These superfields are crucial in order to understand the interplay between (\ref{gamma}) and (\ref{delta}). For example if we replace the fields $q$ and $p$ with the superfields $Q$ and $P$ in the Lagrangian $L$ appearing in the QPI (\ref{delta}) and we integrate over $\theta$ and $\bar{\theta}$ then we obtain, modulo some surface terms, just the $\widetilde{\cal L}$ appearing in the CPI (\ref{gamma}):
\begin{equation}
\displaystyle
i\int d\theta d\bar{\theta}\, L[Q,P]=\widetilde{\cal L} -\frac{d}{dt} (\lambda_p p+i\bar{c}_pc^p). \label{epsilon}
\end{equation}
The surface terms in (\ref{epsilon}) can be removed using, from the beginning, the eigenstates of a complete set of commuting operators different from (\ref{alfa}).  
For example, the operators $(\hat{q},\hat{\lambda}_p,\hat{c}^q,\hat{\bar{c}}_p)$,
which appear in the same multiplet $Q(t,\theta,\bar{\theta})$ of Eq. (\ref{fi}),
make up a complete set of commuting operators.
Their simultaneous eigenstates $|q,\lambda_p,c^q,\bar{c}_p\rangle$ 
satisfy the following eigenvalue equation: $\hat{Q}\,|q,\lambda_p,c^q,\bar{c}_p\rangle=
Q\,|q,\lambda_p,c^q,\bar{c}_p\rangle$. 
Therefore we can identify $|Q\rangle\equiv |q,\lambda_p,c^q,\bar{c}_p\rangle$.
The kernel of propagation between these states 
$\langle Q;t|Q_{\scriptscriptstyle 0};t_{\scriptscriptstyle 0}\rangle$ can be obtained from (\ref{gamma}) via a Fourier transform on the initial and final variables labeled by $p$. This operation cancel exactly the surface terms in (\ref{epsilon}) and changes the path integral (\ref{gamma}) into:
\begin{equation}
\displaystyle 
\langle Q;t|Q_{\scriptscriptstyle 0};t_{\scriptscriptstyle 0}\rangle=\int {\mathscr D}^{\prime \prime}Q{\mathscr D}P\;
\textrm{exp}\biggl[i \int_{t_0}^{t} i d\tau d\theta d\bar{\theta}\, L(Q,P)\biggr], \label{zeta}
\end{equation}
where the functional integration over a superfield means a functional integration 
over all the components of the superfield. Now the CPI (\ref{zeta}) has the same form of the QPI (\ref{delta}) and it can be obtained from (\ref{delta})
by: {\bf 1)} replacing the fields $q$, $p$ with the superfields $Q$, $P$ and {\bf 2)} extending the integration over $\tau$ to an integration over the ``supertime" $(\tau,\theta,\bar{\theta})$ multiplied by $\hbar$, i.e.
$\displaystyle \int d\tau \, \longrightarrow \, i\hbar \int d\tau d\theta d\bar{\theta}$. 
For a detailed analysis of this dequantization procedure we refer the reader to Ref. \cite{deq}.

\section{Spin and Grassmann variables}

The spin one half degrees of freedom of a particle are usually described via a two-dimensional Hilbert space ${\mathcal{H}}_{\scriptscriptstyle S}$
spanned, for example, by the two eigenstates, $|+\rangle$
and $|-\rangle$, of the third component of the spin operator $\hat{S}_z$:
\begin{displaymath}
\displaystyle \hat{S}_z\, |+\rangle =\frac{\hbar}{2}|+\rangle,\qquad\quad \hat{S}_z\,
|-\rangle=-\frac{\hbar}{2}\,|-\rangle.
\end{displaymath}
The most general element of the Hilbert space ${\mathcal{H}}_{\scriptscriptstyle S}$ can then be written
as a linear combination with complex coefficients of the eigenstates above:
\begin{equation}
|\psi\rangle =\psi_{\scriptscriptstyle 0}|+\rangle +\psi_{\scriptscriptstyle 1}\, |-\rangle,\qquad
\qquad \psi_{\scriptscriptstyle 0},\psi_{\scriptscriptstyle 1}\in \mathbb{C}. \label{gra0}
\end{equation}
In the basis $\Bigl\{|+\rangle,|-\rangle\Bigr\}$ we can represent $|\psi\rangle$ as a two-component vector $\displaystyle \begin{pmatrix}
\psi_{\scriptscriptstyle 0}\cr \psi_{\scriptscriptstyle 1}\end{pmatrix}$
and the operator $\hat{S}_z$ as the following diagonal matrix 
$\hat{S}_z=\displaystyle \frac{\hbar}{2}\begin{pmatrix}1 & 0\cr 0 & -1\end{pmatrix}$.

Now we want to prove that there exists an isomorphism between the Hilbert space 
${\mathcal H}_{\scriptscriptstyle S}$ of a particle with spin
and the Hilbert space ${\mathcal H}_{\scriptscriptstyle G}$ that describes a particle with one Grassmannian odd degree of freedom.
This last Hilbert space is characterized by two nilpotent Grassmann operators $\hat{\xi}$
and $\hat{\bar{\xi}}$ that satisfy the 
anticommutator $[\hat{\xi},\hat{\bar{\xi}}]_{\scriptscriptstyle +}=1$ 
and the Hermiticity condition $\hat{\xi}^{\dagger}=\hat{\bar{\xi}}$. Combining $\hat{\xi}$ and $\hat{\bar{\xi}}$ it is possible to build the Hermitian operator 
$\displaystyle \hat{N}=[\hat{\bar{\xi}}\hat{\xi}-\hat{\xi}\hat{\bar{\xi}}]/2$. Since
$\displaystyle \hat{N}^2=1/4$ the only eigenvalues of $\hat{N}$ are $\pm 1/2$ and the associated eigenstates make up a basis for the Hilbert space ${\mathcal H}_{\scriptscriptstyle G}$. If we represent 
$\displaystyle \hat{\xi}$ as the operator of multiplication by $\xi$ 
and $\hat{\bar{\xi}}$ as the derivative operator $\hat{\bar{\xi}}=\frac{\partial}{\partial \xi}$, 
then the eigenstate of $\hat{N}$ with eigenvalue $+1/2$ is simply given by the real number $1$, while the eigenstate of $\hat{N}$ 
with eigenvalue $-1/2$ is the anticommuting number $\xi$. For details see for example Refs. \cite{vH}-\cite{salo}. Since $\{1,\xi\}$ is a basis 
for the Hilbert space ${\mathcal H}_{\scriptscriptstyle G}$, every wave function $\psi$ can be expressed as a linear combination of $1$ and $\xi$ with complex coefficients:
\begin{equation}
\psi(\xi)=\psi_{\scriptscriptstyle 0}+\psi_{\scriptscriptstyle 1}\xi, \qquad\quad 
\psi_{\scriptscriptstyle 0},\psi_{\scriptscriptstyle 1}\in \mathbb{C}. \label{gra}
\end{equation}
Eq. (\ref{gra}) is nothing else than the Taylor expansion of the most 
general function $\psi(\xi)$ of the Grassmann
variable $\xi$. At this point it should be clear that there is an isomorphism between the $\psi(\xi)$ in (\ref{gra}) and the wave functions (\ref{gra0}) that usually describe a particle with spin. This isomorphism among states implies also an isomorphism among operators. In fact if we represent $\psi(\xi)$ as a two-component vector $\displaystyle \begin{pmatrix}\psi_{\scriptscriptstyle 0}\cr
\psi_{\scriptscriptstyle 1}\end{pmatrix}$ then we have that $\displaystyle \hat{N}=\frac{1}{2}\begin{pmatrix}1 & 0 \cr
0 & -1\end{pmatrix}$. Therefore $\hat{N}$ acts, modulo the factor $\hbar$, just as the third component of the spin operator and we can identify $\hat{S}_z=\hbar \hat{N}$. Using the isomorphism between (\ref{gra0}) and (\ref{gra}), we can associate the following Grassmann operators with the other two components of $\hat{\bf S}$:
\begin{equation}
\displaystyle \hat{S}_x=\frac{\hbar}{2}\begin{pmatrix}0 & 1\cr 1 & 0\end{pmatrix}=
\frac{\hbar}{2}(\hat{\bar{\xi}}+\hat{\xi}),
\qquad\qquad \hat{S}_y=\frac{\hbar}{2}\begin{pmatrix}0 & -i\cr i & 0\end{pmatrix}=
\frac{i\hbar}{2}(\hat{\xi}-\hat{\bar{\xi}}). \label{alreesse}
\end{equation}

So every operator depending on $\hat{\bf S}$ can be expressed 
as a Grassmann operator acting on the wave functions $\psi(\xi)$. For example, 
the Hamiltonian describing the interaction of a spinning particle with a constant magnetic
field, $\displaystyle \hat{H}=-\frac{e}{mc}{\bf B}\cdot \hat{\bf S}$, can be rewritten in terms of Grassmann
operators as:
\begin{eqnarray}
\hat{H}&\hspace{-0.2cm}=&\hspace{-0.2cm}-\frac{e}{mc}\biggl[B_x\frac{\hbar}{2}(\hat{\bar{\xi}}+\hat{\xi})+
B_yi\frac{\hbar}{2}
(\hat{\xi}-\hat{\bar{\xi}})+B_z\frac{\hbar}{2}(\hat{\bar{\xi}}\hat{\xi}-\hat{\xi}\hat{\bar{\xi}})\biggr]=
\nonumber \medskip\\
&\hspace{-0.2cm}=&\hspace{-0.2cm} -\mu_{\scriptscriptstyle B}
\biggl[B_z+(B_x+iB_y)\hat{\xi}+(B_x-iB_y)\hat{\bar{\xi}}-
2B_z\hat{\xi}\hat{\bar{\xi}}\biggr], \label{op}
\end{eqnarray}
where we have indicated with $\displaystyle \mu_{\scriptscriptstyle B}=\frac{e\hbar}{2mc}$ the Bohr magneton. 

The action of the operator (\ref{op}) on a generic wave function $\psi(\xi)$ can be written
in the following two ways:
\begin{eqnarray}
\displaystyle \hat{H}\psi(\xi) \label{numera}
&=&\int d\xi^{\prime}\widetilde{H}(\xi,\xi^{\prime})\psi(\xi^{\prime}) \nonumber \\
&=&\int d\xi^{\prime} d\bar{\xi}\; \overline{H}(\xi,\bar{\xi})
e^{\bar{\xi}(\xi^{\prime}-\xi)}\psi(\xi^{\prime}),
\end{eqnarray}
where the explicit expressions of the {\it integral kernel} $\widetilde{H}(\xi,\xi^{\prime})$
and of the {\it ordered symbol} $\overline{H}(\xi,\bar{\xi})$ are the following ones \cite{vH}:
\begin{equation}
\begin{array}{l}
\displaystyle \widetilde{H}(\xi,\xi^{\prime})=-\mu_{\scriptscriptstyle B}(B_x-iB_y)
-\mu_{\scriptscriptstyle B}B_z\xi-\mu_{\scriptscriptstyle B}B_z\xi^{\prime}+\mu_{\scriptscriptstyle B}
(B_x+iB_y)\xi\xi^{\prime} \medskip \\
\displaystyle \overline{H}(\xi,\bar{\xi})=-\mu_{\scriptscriptstyle B}B_z-\mu_{\scriptscriptstyle B}
(B_x+iB_y)\xi-\mu_{\scriptscriptstyle B}(B_x-iB_y)\bar{\xi}
+2\mu_{\scriptscriptstyle B}B_z\xi \bar{\xi}. \label{ham}
\end{array}
\end{equation}
Let us remember that the evolution operator $\displaystyle \hat{U}(t)=e^{-i t\hat{H}/ \hbar}$ satisfies the property:
\begin{equation}
\displaystyle \hat{U}(t-t_{\scriptscriptstyle 0})=\hat{U}(t-t^{\prime})\hat{U}(t^{\prime}-t_{\scriptscriptstyle 0}), \label{evolutor}
\end{equation}
so the ordered symbol associated with the LHS of (\ref{evolutor}) 
must be given by the ordered symbol of the product of the two operators appearing on the RHS, i.e.:
\begin{equation}
\displaystyle \overline{U}(\xi,t;\bar{\xi}_{\scriptscriptstyle 0}, t_{\scriptscriptstyle 0})=\int d\xi^{\prime}d\bar{\xi}^{\prime} \, e^{(\bar{\xi}^{\, \prime}-\bar{\xi}_{\scriptscriptstyle 0})(\xi^{\prime}-\xi)} \,
\overline{U}(\xi,t;\bar{\xi}^{\prime},t^{\prime})
\overline{U}(\xi^{\prime},t^{\prime};\bar{\xi}_{\scriptscriptstyle 0},t_{\scriptscriptstyle 0}). 
\label{13-2}
\end{equation}
Let us consider the time interval $(t_{\scriptscriptstyle 0},t)$ and divide it into $N+1$ steps of length $\epsilon$. Then $t-t_{\scriptscriptstyle 0}=(N+1)\epsilon$ and $\hat{U}(t-t_{\scriptscriptstyle 0})=[\hat{U}(\epsilon)]^{N+1}$. Applying Eq. (\ref{13-2}) it is possible to derive, in the limit $N\to \infty$ and $\epsilon \to 0$, the following expression \cite{vH}:
\begin{displaymath}
\displaystyle \overline{U}(\xi,t;\bar{\xi}_{\scriptscriptstyle 0},t_{\scriptscriptstyle 0})=\lim_{N\to \infty} \biggl\{e^ 
{\bar{\xi}_{\scriptscriptstyle 0}(\xi_{\scriptscriptstyle N+1}-\xi_{\scriptscriptstyle 0})}
\int \prod_{k=1}^N[d\xi_kd\bar{\xi}_k]\exp \biggl[\frac{i\epsilon}{\hbar}
\sum_{l=0}^N \biggl(i\hbar\bar{\xi}_l\frac{\xi_{l+1}-\xi_l}{\epsilon}-
\overline{H}(\xi_{l+1},\bar{\xi_l})\biggr)\biggr]\biggr\},
\end{displaymath}
where $\xi$ has to be identified with $\xi_{\scriptscriptstyle N+1}$. From the expression of $\overline{U}$ and Eq. (\ref{numera}) we can derive the following expression for
the integral kernel $\widetilde{U}$:
\begin{displaymath}
\displaystyle \widetilde{U}(\xi,t;\xi_{\scriptscriptstyle 0},t_{\scriptscriptstyle 0})=
\lim_{N\to \infty} \int d\bar{\xi}_{\scriptscriptstyle 0}\int \prod_{k=1}^N[d\xi_kd\bar{\xi}_k]
\exp \biggl[\frac{i\epsilon}{\hbar} \sum_{l=0}^N\biggl(i\hbar\bar{\xi_l}\frac{\xi_{l+1}-\xi_l}{\epsilon}
-\overline{H}(\xi_{l+1},\bar{\xi}_l)\biggr)\biggr]. \label{16-2}
\end{displaymath}
The kernel of evolution can be written in the following path integral form:
\begin{equation}
\displaystyle \widetilde{U}(\xi, t;\xi_{\scriptscriptstyle 0},t_{\scriptscriptstyle 0})=\int {\mathscr D}^{\prime\prime}\xi{\mathscr D}\bar{\xi} \exp \biggl[\frac{i}{\hbar}\int_{t_0}^t
d\tau [i\hbar \bar{\xi}\dot{\xi}-\overline{H}(\xi,\bar{\xi})]\biggr]. \label{16-3}
\end{equation}
The $ \widetilde{U}$ above propagates the wave functions $\psi(\xi)=\psi_{\scriptscriptstyle 0}+
\psi_{\scriptscriptstyle 1}\xi$ according to the equation:
\begin{displaymath}
\psi(\xi,t)=\int d\xi_{\scriptscriptstyle 0}\, \widetilde{U}(\xi,t;\xi_{\scriptscriptstyle 0},t_{\scriptscriptstyle 0})\,\psi(\xi_{\scriptscriptstyle 0},t_{\scriptscriptstyle 0}),
\end{displaymath}
which is completely equivalent to the Pauli equation for the spin part of a quantum wave function \cite{messiah}:
\begin{displaymath}
\displaystyle i\hbar \frac{\partial}{\partial t}\begin{pmatrix}\psi_{\scriptscriptstyle 0}\cr
\psi_{\scriptscriptstyle 1}\end{pmatrix}=\hat{H}_{\scriptscriptstyle P}\begin{pmatrix}
\psi_{\scriptscriptstyle 0} \cr 
\psi_{\scriptscriptstyle 1}\end{pmatrix}, \qquad \quad \hat{H}_{\scriptscriptstyle P}
=-\mu_{\scriptscriptstyle B} \begin{pmatrix}B_z & B_x-iB_y \cr
B_x+iB_y & -B_z\end{pmatrix}.
\end{displaymath}

\section{Grassmannian Classical Path Integral for Spinning Particles}

In this section we want to build the CPI that lies behind the Grassmannian QPI for spin degrees of freedom given by Eq. (\ref{16-3}). First of all let us align the magnetic field with the $z$ axis. In this case the Hamiltonian $\overline{H}(\xi,\bar{\xi})$ of Eq. (\ref{ham}) becomes a Grassmannian even object and the path integral (\ref{16-3}) reduces to:
\begin{equation}
\displaystyle \langle \xi; t| \xi_{\scriptscriptstyle 0};t_{\scriptscriptstyle 0} \rangle \equiv \widetilde{U}(\xi, t;\xi_{\scriptscriptstyle 0},t_{\scriptscriptstyle 0})=\int {\mathscr D}^{\prime\prime}\xi{\mathscr D}\bar{\xi}\; \exp \biggl[i\int_{t_0}^t
d\tau L(\xi,\bar{\xi})\biggr] \label{qpath}
\end{equation}
with the Lagrangian $L(\xi,\bar{\xi})$ given by:
\begin{equation}
\displaystyle L(\xi,\bar{\xi})=i\bar{\xi}\dot{\xi}+\frac{eB}{2mc}(1-2\xi\bar{\xi}).  \label{mi}
\end{equation}
From this Lagrangian we can derive the following Euler-Lagrange equation of motion:
\begin{equation}
\displaystyle \dot{\xi}-\frac{ieB}{mc}\xi=0,\qquad \;\; \dot{\bar{\xi}}+\frac{ieB}{mc}\bar{\xi}=0.
\label{classeq}
\end{equation}
Starting from these Grassmannian odd equations of motion and following steps similar to the ones analyzed in {\bf Sec. 2}, we can derive the associated CPI:
\begin{displaymath}
\displaystyle \langle \xi,\bar{\xi};t|\xi_{\scriptscriptstyle 0},\bar{\xi}_{\scriptscriptstyle 0};t_{\scriptscriptstyle 0}\rangle=
\int {\mathscr D}^{\prime\prime}\xi{\mathscr D}^{\prime\prime}\bar{\xi} \;\widetilde{\delta}\Bigl[\xi-
\xi_{\textrm{cl}}(t;\xi_{\scriptscriptstyle 0},t_{\scriptscriptstyle 0})\Bigr]\, \widetilde{\delta}\Bigl[\bar{\xi}-\bar{\xi}_{\textrm{cl}}(t;\bar{\xi}_{\scriptscriptstyle 0},t_{\scriptscriptstyle 0})\Bigr]. 
\end{displaymath}
We can then pass from the delta of the solutions to the delta of the equations of motion, as follows:
\begin{equation}
\begin{array}{l}
\qquad \qquad  \displaystyle \langle \xi,\bar{\xi};t|\xi_{\scriptscriptstyle 0},\bar{\xi}_{\scriptscriptstyle 0};t_{\scriptscriptstyle 0}\rangle =  \label{inverse} \medskip \\
=\displaystyle \int {\mathscr D}^{\prime\prime}\xi{\mathscr D}^{\prime\prime}\bar{\xi}\;\;
\widetilde{\delta}\biggl(\dot{\xi}-\frac{ieB}{mc}\xi\biggr)
\widetilde{\delta}\biggl(\dot{\bar{\xi}}+\frac{ieB}{mc}\bar{\xi}\biggr)
\textrm{det}^{-1}\begin{pmatrix}\partial_t-\frac{ieB}{mc} & 0 \cr 0 &
\partial_t+\frac{ieB}{mc}\end{pmatrix}. 
\end{array}
\end{equation}
Since the phase space variables $\phi^a\equiv (\xi,\bar{\xi})$ are 
Grassmannian odd, in Eq. (\ref{inverse})
there appears the inverse of a determinant instead of the determinant of Eq. (\ref{due}). We can then
exponentiate the functional Dirac delta of the Grassmannian odd equations of motion $\widetilde{\delta}(A)$ via the Grasmmannian odd variables $\lambda_a \equiv (\lambda_{\xi},\lambda_{\bar{\xi}})$ and the inverse of the functional determinant 
$D$ via the Grassmmannian even auxiliary variables $c^a \equiv (c^{\xi},c^{\bar{\xi}})$ and $\bar{c}_a \equiv (\bar{c}_{\xi},\bar{c}_{\bar{\xi}})$ according to the following equations:
\begin{equation}
\displaystyle \widetilde{\delta}(A)=\int {\mathscr D}\lambda \exp \biggl[i\int d\tau \, \lambda A\biggr],
\qquad  \textrm{det}^{-1}D=\int {\mathscr D}c {\mathscr D} \bar{c} \exp \biggl[i\int d\tau \,i\bar{c}_a D^a_b c^b\biggr]. \label{premessa}
\end{equation}
Using the expression (\ref{premessa}) into (\ref{inverse}), the classical kernel of propagation becomes:
\begin{displaymath}
\displaystyle \langle \phi;t|\phi_{\scriptscriptstyle 0};t_{\scriptscriptstyle 0}\rangle
=\int {\mathscr D}^{\prime\prime}\phi {\mathscr D}\lambda
{\mathscr D}c {\mathscr D}\bar{c} \,\exp \biggl[i\int_{t_0}^t d\tau \widetilde{\cal L}\biggr],
\end{displaymath}
where $\widetilde{\cal L}$ is the following Lagrangian:
\begin{equation}
\begin{array}{l} 
\displaystyle \widetilde{\cal L} = \lambda_{\xi}\dot{\xi}+\lambda_{\bar{\xi}}\dot{\bar{\xi}}
+i\bar{c}_{\xi}\dot{c}^{\xi}+i\bar{c}_{\bar{\xi}}\dot{c}^{\bar{\xi}}-\widetilde{\cal H}
\label{tildezeta}  \bigskip \\
 \displaystyle \widetilde{\cal H} = \frac{ieB}{mc}\Bigl(\lambda_{\xi}\xi-
\lambda_{\bar{\xi}}\bar{\xi}+i\bar{c}_{\xi}
c^{\xi}-i\bar{c}_{\bar{\xi}}c^{\bar{\xi}}\Bigr).
\end{array}
\end{equation}
From the kinetic terms of the Lagrangian (\ref{tildezeta}) we can deduce that the only graded commutators different from zero are:
\begin{equation}
\displaystyle [\xi,\lambda_{\xi}]=i, \quad [\bar{\xi},\lambda_{\bar{\xi}}]=i, \qquad
[c^{\xi},\bar{c}_{\xi}]=1, \quad [c^{\bar{\xi}},\bar{c}_{\bar{\xi}}]=1. \label{elettra}
\end{equation}
Since the operators $\hat{\phi}^a=(\hat{\xi},\hat{\bar{\xi}})$ commute with the operators $\hat{c}^a=(\hat{c}^{\xi},\hat{c}^{\bar{\xi}})$, it is appropriate to consider the kernel of
propagation in the $(\phi,c)$-space:
\begin{equation}
\langle \phi,c;t|\phi_{\scriptscriptstyle 0},c_{\scriptscriptstyle 0};t_{\scriptscriptstyle 0}\rangle=
\int {\mathscr D}^{\prime\prime}\phi {\mathscr D}\lambda {\mathscr D}^{\prime\prime}c
{\mathscr D}\bar{c}\; \textrm{exp}\biggl[i
\int_{t_0}^{t} d\tau\,\widetilde{\cal L}\biggr]. \label{fraca}
\end{equation}
The graded commutators (\ref{elettra}) can be realized by considering
$\hat{\xi}$, $\hat{\bar{\xi}}$, $\hat{c}^{\xi}$ and $\hat{c}^{\bar{\xi}}$ as operators of multiplication and $\hat{\lambda}_{\xi}$, $\hat{\lambda}_{\bar{\xi}}$, $\hat{\bar{c}}_{\xi}$ and $\hat{\bar{c}}_{\bar{\xi}}$ as derivative operators:
\begin{displaymath}
\displaystyle \hat{\lambda}_{\xi}=i\frac{\partial}{\partial \xi}, \qquad
\hat{\lambda}_{\bar{\xi}}=i\frac{\partial}{\partial \bar{\xi}} , \qquad 
\hat{\bar{c}}_{\xi}=-\frac{\partial}{\partial c^{\xi}},
\qquad \hat{\bar{c}}_{\bar{\xi}}=-\frac{\partial}{\partial c^{\bar{\xi}}}.
\end{displaymath}
Basically the kernel of propagation (\ref{fraca}) generates the evolution of the wave functions $\psi$ according to the equation of motion
\begin{equation}
\displaystyle i\frac{\partial}{\partial t}\psi(\phi,c)=
\hat{\widetilde{\cal H}}\psi(\phi,c), \label{blimunda}
\end{equation}
where $\hat{\widetilde{\cal H}}$ is the operator associated to the Hamiltonian of Eq. (\ref{tildezeta}):
\begin{displaymath}
\displaystyle \hat{\widetilde{\cal H}}=-\frac{eB}{mc}
\Bigl(\frac{\partial}{\partial \xi}\xi-
\frac{\partial}{\partial \bar{\xi}}\bar{\xi}-\frac{\partial}{\partial c^{\xi}}c^{\xi}
+\frac{\partial}{\partial c^{\bar{\xi}}}c^{\bar{\xi}}
\Bigr).
\end{displaymath}
We know \cite{gozzi} that the CPI (\ref{gamma}) reproduces the kernel of evolution associated with a generalized Liouville equation for classical statistical mechanics. Analogously Eq. (\ref{blimunda}), which lies behind the path integral (\ref{fraca}), can be considered as a sort of classical Liouville equation for a spinning particle.

Is it possible to connect the QPI (\ref{qpath}) and the CPI (\ref{fraca}) via the superfield procedure described in {\bf Sec. 2}? 
The answer is: yes, provided we give the following definition of the superfields:
\begin{equation}
\displaystyle \Xi=\xi+\theta
c^{\xi}-i\bar{\theta}\bar{c}_{\bar{\xi}}-\bar{\theta}\theta\lambda_{\bar{\xi}}, \qquad\quad
\bar{\Xi}=\bar{\xi}+\theta c^{\bar{\xi}}-i\bar{\theta}\bar{c}_{\xi}
-\bar{\theta}\theta\lambda_{\xi}. \label{fersup}
\end{equation}
With this definition we can easily pass from the Lagrangian $L$ of Eq. (\ref{mi}) to the 
Lagrangian $\widetilde{\cal L}$ of Eq. (\ref{tildezeta}), replacing the fields $\xi$ and $\bar{\xi}$ with the superfields $\Xi$ and $\bar{\Xi}$ of Eq. (\ref{fersup}) and integrating in $\theta$ and $\bar{\theta}$:
\begin{equation}
\displaystyle i\int d\theta d\bar{\theta} \, L(\Xi,\bar{\Xi})=
\widetilde{\cal L}-\frac{d}{dt}(\lambda_{\bar{\xi}}\bar{\xi}+i\bar{c}_{\bar{\xi}}c^{\bar{\xi}}).
\label{surf}
\end{equation}
The surface terms appearing in (\ref{surf}) involve the variables $\bar{\xi}$,
$\lambda_{\bar{\xi}}$, $c^{\bar{\xi}}$ and $\bar{c}_{\bar{\xi}}$, and they can be reabsorbed, as in the bosonic case analyzed in {\bf Sec. 2}, via a partial 
Fourier transform with respect to the variables $(\bar{\xi},\lambda_{\bar{\xi}})$ and 
$(c^{\bar{\xi}},\bar{c}_{\bar{\xi}})$ respectively. This means that if we change the representation and we consider the kernel of propagation between the eigenstates of the superfield $\hat{\Xi}$, which are $|\Xi \rangle=|\xi,\lambda_{\bar{\xi}},c^{\xi},\bar{c}_{\bar{\xi}}\rangle$, we get the following path integral:
\begin{equation}
\displaystyle \langle \Xi;t|
\Xi_{\scriptscriptstyle 0};t_{\scriptscriptstyle 0}\rangle=
\int {\mathscr D}^{\prime\prime}\Xi \, {\mathscr D}\bar{\Xi}\;
\textrm{exp}\biggl[i\int_{t_0}^t id\tau d\theta d\bar{\theta}\,L(\Xi,\bar{\Xi})\biggr],
\label{nome}
\end{equation}
where the functional measure is given by:
\begin{displaymath}
\displaystyle {\mathscr D}^{\prime\prime}\Xi\equiv {\mathscr D}^{\prime\prime}
\xi {\mathscr D}^{\prime\prime}\lambda_{\bar{\xi}}{\mathscr D}^{\prime\prime} c^{\xi}{\mathscr D}^{\prime\prime} \bar{c}_{\bar{\xi}}, \qquad 
{\mathscr D}\bar{\Xi}\equiv {\mathscr D}\bar{\xi}{\mathscr D}\lambda_{\xi}{\mathscr D}c^{\bar{\xi}}{\mathscr D}\bar{c}_{\xi}.
\end{displaymath}

This means that the same dequantization procedure analyzed in {\bf Sec. 2} works also in the case of particles with spin analyzed above: to go from the quantum path integral (\ref{qpath}) to the classical one (\ref{nome}) we must replace everywhere the fields $\xi$ and $\bar{\xi}$ with the superfields $\Xi$ and $\bar{\Xi}$ of Eq. (\ref{fersup}) and extend the integration from time to supertime $\int d\tau \, \rightarrow \, i \int d\tau d\theta d\bar{\theta}$. 
Before concluding this section, we should point out that a generalization of the CPI including Grassmann variables was proposed first in Ref. \cite{marnelius}.

\section{Bosonic Classical Path Integral for Spinning Particles}

Another possibility to implement a path integral for the spinning particle in quantum mechanics is based on the coadjoint orbit method \cite{arnold}. There is a theorem which says that
{\it Every orbit of the coadjoint action of a Lie group possesses a symplectic structure}, see the last of Refs. \cite{pauri}. 
In the case of the group SO(3) the coadjoint orbits can be identified with the spheres $S^2$ and they are parameterized by their radius \cite{alek}. If we use as coordinates $x^{\scriptscriptstyle 1}$, $x^{\scriptscriptstyle 2}$ and $x^{\scriptscriptstyle 3}$ satisfying $\sum_i (x^i)^2=\lambda^2$, then the symplectic form on the two-sphere $S^2$ of radius $\lambda$ is given by $\displaystyle \Omega=\frac{1}{2\lambda^2} \epsilon^{\alpha\beta\gamma}x^{\alpha}dx^{\beta} dx^{\gamma}$, where $\epsilon^{\alpha\beta\gamma}$ are the structure constants of the group itself:
$\{x^{\alpha},x^{\beta}\}_{\scriptscriptstyle P}=\epsilon^{\alpha \beta \gamma} x^{\gamma}$. The Darboux variables are given by the spherical coordinates \cite{alek}:
\begin{equation}
\displaystyle  x^{\scriptscriptstyle 1}=
\lambda\sin\theta\cos\varphi,\qquad x^{\scriptscriptstyle 2}=\lambda\sin\theta\sin\varphi,\qquad
x^{\scriptscriptstyle 3}=\lambda \cos\theta  \label{coord}
\end{equation}
and the symplectic form can be written as $\Omega=\lambda \,d\varphi \,d\sin \theta$.
The one-form $\omega=-d^{-1}\Omega$ entering the definition of the action can be identified with 
\begin{equation}
\omega=(\gamma+\lambda \cos \theta)d\varphi, \label{symplectic}
\end{equation}
while the associated action becomes $S=\int \omega$. This is just the form of the action considered in \cite{nielsen} and \cite{johnson}. 

More precisely, taking into account also the interaction with an external magnetic field $B$ pointing along the $z$ axis, an appropriate Lagrangian to describe a classical
action that fixes ``{\it the magnitude of the spin, leaving its direction free}" \cite{nielsen} 
is given by:
\begin{equation}
\displaystyle L(\varphi,\theta)=(\gamma+\lambda \cos \theta) \dot{\varphi}+\lambda\mu B
\cos \theta. \label{ttilde}
\end{equation}
Since the constant term $\gamma$ in (\ref{ttilde}) does not play any dynamical role and does not enter the classical equations of motion, from now on we will disregard it in the implementation of the CPI.
The classical equations of motion that can be derived from the Lagrangian (\ref{ttilde}) 
are equivalent to the following ones:
\begin{equation}
\displaystyle \frac{d}{dt}[\lambda\cos\theta(t)]=0, \qquad
(\dot{\varphi}(t)+\mu B) \sin \theta(t)=0 \label{eqq}
\end{equation}
whose solutions are given by:
\begin{equation}
\displaystyle \theta(t)=\theta_{\scriptscriptstyle 0}=\textrm{const.}, \qquad 
\varphi(t)=\varphi_{\scriptscriptstyle 0}-\mu B(t-t_{\scriptscriptstyle 0}). \label{eqq2}
\end{equation}
Since $\theta$ is constant and $\varphi$ varies linearly with time, the particle
describes a circumference that is the contour of the basis of a cone. 
The classical motion of the particle turns out to be a precession in the magnetic field. Such a motion is periodic with period given by $\displaystyle T=\frac{2\pi}{\mu B}$.

Let us now write down the equations of motion in a Hamiltonian form. First of all,
from the Lagrangian (\ref{ttilde}) and the definition itself of conjugate momenta, we can derive the following primary constraints:
\begin{displaymath}
\displaystyle \Phi_{\scriptscriptstyle 1}: p_{\theta}=0, \qquad \quad \Phi_{\scriptscriptstyle 2}: p_{\varphi}-\lambda\cos \theta =0.
\end{displaymath}
Implementing the Dirac procedure we have that the total Hamiltonian is given by:
\begin{displaymath}
\displaystyle H_{\scriptscriptstyle T}=p_{\theta}\dot{\theta}+p_{\varphi}\dot{\varphi}
-\lambda \cos \theta \, \dot{\varphi} -\lambda \mu B\cos \theta
+v_{\scriptscriptstyle 1}\Phi_{\scriptscriptstyle 1}+v_{\scriptscriptstyle 2}\Phi_{\scriptscriptstyle 2}.
\end{displaymath} 
If we impose that the constraints are conserved in time we can then determine the Lagrangian multipliers $v_{\scriptscriptstyle 1}$ and $v_{\scriptscriptstyle 2}$. Doing so the total Hamiltonian turns out to be:
\begin{equation}
H=-\lambda\mu B \cos \theta. \label{total}
\end{equation}
The Poisson brackets among the constraints of the theory are $\{\Phi_{\scriptscriptstyle 1}, \Phi_{\scriptscriptstyle 2}\}_{\scriptscriptstyle P}=-\lambda \sin \theta$, so the matrix entering the definition of the Dirac brackets is:
\begin{displaymath}
\displaystyle C_{ab}=\{\Phi_a,\Phi_b\}_{\scriptscriptstyle P}^{-1}=\begin{pmatrix} 0 & \frac{1}{\lambda\sin \theta}\cr
-\frac{1}{\lambda \sin\theta} & 0\end{pmatrix}.
\end{displaymath}
The only non-zero Dirac brackets are the ones between $\varphi$ and $\lambda\cos \theta$:
\begin{equation}
\displaystyle \{\varphi, \lambda\cos \theta\}_{\scriptscriptstyle D}=
\{\varphi, \lambda \cos \theta\}_{\scriptscriptstyle P}-\{\varphi, \Phi_a\}_{\scriptscriptstyle P}\,
C_{ab}\, \{\Phi_b,\lambda\cos \theta \}_{\scriptscriptstyle P}=1. \label{dirac}
\end{equation}
It is in this sense that we can consider $\varphi$ and $\eta=\lambda\cos \theta$ as
canonically conjugated variables. It should be clear that $\varphi$ and $\eta$ are canonical coordinates just as a consequence of the particular form of the action and, consequently, of the symplectic structure (\ref{symplectic}) associated with the coadjoint orbits of the group SO(3). If we introduce a unique variable $\phi^a=(\varphi,\eta)$, with $a=1,2$, then we can write the equations of motion in terms of the total Hamiltonian (\ref{total}) and of the Dirac brackets (\ref{dirac}) as $\dot{\phi}^a=\{\phi^a,H\}_{\scriptscriptstyle D}$ or, introducing the matrix $\omega^{ab}=\begin{pmatrix}0 & 1 \cr -1 & 0\end{pmatrix}$, as $\dot{\phi}^a=\omega^{ab}\partial_bH$. 

The CPI can be easily realized following the same steps reviewed in {\bf Sec. 2}. From (\ref{eqq2}) we derive that the functional Dirac delta of the solutions of the equations of motion becomes:
\begin{equation}
\displaystyle K(\varphi, \eta;t|\varphi_{\scriptscriptstyle 0},
\eta_{\scriptscriptstyle 0}; t_{\scriptscriptstyle 0})= \int {\mathscr D}^{\prime\prime}\varphi {\mathscr D}^{\prime\prime}\eta \; \widetilde{\delta} (\eta-\eta_{\scriptscriptstyle 0})\widetilde{\delta}(\varphi-\varphi_{\scriptscriptstyle 0}+\mu B \left( t-t_{\scriptscriptstyle 0}) \right). \label{clker}
\end{equation}
In terms of the Dirac delta of the equations of motion the kernel of propagation 
(\ref{clker}) can be rewritten as:
\begin{displaymath}
\displaystyle K(\phi^a; t|\phi^a_{\scriptscriptstyle 0};t_{\scriptscriptstyle 0})
=\int {\mathscr D}^{\prime\prime}\phi \;\widetilde{\delta}(\dot{\phi}^a-\omega^{ab}\partial_bH)\,
\textrm{det}[\delta_b^a \partial_t -\omega^{ac}\partial_c\partial_bH],
\end{displaymath}
which, repeating the same steps analyzed in {\bf Sec. 2}, produces the following standard expression for the classical kernel of propagation:
\begin{equation}
\displaystyle \langle \phi,c;t|\phi_{\scriptscriptstyle 0},c_{\scriptscriptstyle 0};t_{\scriptscriptstyle 0}\rangle=\int {\mathscr D}^{\prime\prime}\phi^a {\mathscr D}\Lambda_a {\mathscr D}^{\prime\prime} c^a {\mathscr D}\bar{c}_a \; \exp i\int_{t_0}^t d\tau \widetilde{\cal L}, \label{cpi}
\end{equation}
where we have used $\Lambda$ instead of $\lambda$ to avoid confusion with the radius of the $S^2$ sphere, while $\widetilde{\cal L}$ is the following Lagrangian:
\begin{equation}
\displaystyle \widetilde{\cal L}=\Lambda_a\dot{\phi}^a+i\bar{c}_a\dot{c}^a-
\widetilde{\cal H}, \qquad \quad \widetilde{\cal H}=\Lambda_a\omega^{ab}\partial_bH+i\bar{c}_a\omega^{ab}\partial_b\partial_dHc^d. \label{eqquat}
\end{equation}
Because of the particular form of the Hamiltonian $H$ of Eq. (\ref{total}), the $\widetilde{\cal H}$ in (\ref{eqquat}) lacks the term with the double derivative and reduces to the following Liouvillian:
\begin{displaymath}
\widetilde{\cal H}=\Lambda_a\omega^{ab}\partial_bH=\Lambda_{\varphi}\omega^{\varphi\eta}\partial_{\eta}H=-\mu B\Lambda_{\varphi}.
\end{displaymath}
The fundamental commutator is $[\varphi,\Lambda_{\varphi}]=i$, so we can represent 
$\displaystyle \hat{\Lambda}_{\varphi}$ as a derivative operator:
$\displaystyle \hat{\Lambda}_{\varphi}=-i\frac{\partial}{\partial \varphi}$. Therefore the 
operator $\hat{\widetilde{\cal H}}$ simply generates a rotation in $\varphi$, like it should be clear from Eq. (\ref{eqq2}). Let us notice that the variation of the Lagrangian (\ref{eqquat}) with respect to $\Lambda_a$ gives the equations of motion we started from, i.e. $\displaystyle \dot{\eta}=0$ and $\dot{\varphi}+\mu B=0$.
The variation with respect to $\bar{c}_a$ gives instead the following equations: $\dot{c}^{\eta}=0$ and $\dot{c}^{\varphi}=0$,
which imply that the length of the Jacobi fields does not increase with time (for the interpretation of the variables $c$ as Jacobi fields, the reader can consult Refs. \cite{gozzi}-\cite{hilbert}). This is consistent with the fact that, varying the initial conditions in $\theta$ (or $\eta$) and $\varphi$, the classical trajectories are given by a series of circumferences with center on the axis of a cone.

The quantum kernel of propagation can instead be written as an integral over $\varphi$ and $\eta$ of the Lagrangian (\ref{ttilde}):
\begin{equation}
\displaystyle \langle \varphi;t|\varphi_{\scriptscriptstyle 0};t_{\scriptscriptstyle 0} \rangle =
\int {\mathscr D}^{\prime\prime} \varphi {\mathscr D}\eta \,\exp\left[\frac{i}{\hbar}\int_{t_0}^t d\tau L(\varphi,\eta)\right], \qquad L(\varphi,\eta)=(\gamma+\eta)\dot{\varphi}+\mu B \eta. \label{behind}
\end{equation}
We refer the reader to the original papers \cite{nielsen}-\cite{johnson} to appreciate the subtleties hidden behind the functional measure $\int {\mathscr D}^{\prime\prime} \varphi {\mathscr D}\eta$ and the role of the term $\gamma$. 
Now the question we want to answer is: how can we connect the classical path integral (\ref{cpi}) and the quantum one (\ref{behind})?

Since the formal structure of the theory is the usual one, we expect that also the definition of the superfields will be the one of Eq. (\ref{fi}), which in this particular case becomes:
\begin{equation}
\left\{
\begin{array}{l}
\displaystyle \widetilde{\varphi}=\varphi+\chi c^{\varphi} +\bar{\chi}\bar{c}_{\eta}+i\bar{\chi}{\chi}
\Lambda_{\eta}, \medskip \\
\displaystyle \widetilde{\eta}=\eta +\chi c^{\eta}-\bar{\chi}\bar{c}_{\varphi}-i\bar{\chi}{\chi}\Lambda_{\varphi}.  \label{deff}
\end{array}
\right.
\end{equation}
We have preferred to change the notation for the superpartners of time from $(\theta, \bar{\theta})$ to $(\chi, \bar{\chi})$, to avoid confusion with the angular variable $\theta$. Now, let us disregard for the moment the constant $\gamma$ in (\ref{behind}), like we have done in the implementation of the CPI. With the definition (\ref{deff}) of the superfields it is possible to reconstruct the Liouvillian 
$\widetilde{\cal H}=-\mu B\Lambda_{\varphi}$ starting from the Hamiltonian
$H=-\mu B\eta$, by simply replacing the fields with the superfields and integrating
the result over $\chi$ and $\bar{\chi}$. In fact:
\begin{displaymath}
\displaystyle i\int d\chi d\bar{\chi} \, H(\widetilde{\varphi},\widetilde{\eta})=-i\mu B\int d\chi d\bar{\chi}
\,\widetilde{\eta}=-\mu B\Lambda_{\varphi}=\widetilde{\cal H}.
\end{displaymath}
Applying the same procedure to the 
Lagrangian of Eq. (\ref{behind}), but with $\gamma=0$, we get the relation:
\begin{equation} 
i\int d\chi d\bar{\chi} \, L\Bigl|_{\gamma=0}[\widetilde{\varphi},\widetilde{\eta}]=\widetilde{\cal L}-\frac{d}{dt} [\Lambda_{\eta}\eta+i\bar{c}_{\eta}c^{\eta}]. \label{derterms}
\end{equation}
As in the cases analyzed in the previous sections, the surface terms $\displaystyle \frac{d}{dt}( \Lambda_{\eta}\eta+i\bar{c}_{\eta}c^{\eta})$ in (\ref{derterms}) can be reabsorbed via the partial Fourier tranforms $\eta\leftrightarrow \Lambda_{\eta}$ and $c^{\eta}\leftrightarrow \bar{c}_{\eta}$ on the initial and final variables. These Fourier transforms turn Eq. (\ref{cpi}) into the kernel of propagation between the states $|\widetilde{\varphi}_{\scriptscriptstyle 0};t_{\scriptscriptstyle 0}\rangle$ and $|\widetilde{\varphi};t\rangle$, where $|\widetilde{\varphi}\,\rangle$ stands for $|\varphi, c^{\varphi}, \bar{c}_{\eta},\Lambda_{\eta}\rangle$. 
This kernel can be written in terms of the superfields (\ref{deff}) as:
\begin{displaymath}
\displaystyle \langle \widetilde{\varphi}; t|\widetilde{\varphi}_{\scriptscriptstyle 0};t_{\scriptscriptstyle 0}\rangle =\int {\mathscr D}^{\prime\prime}\widetilde{\varphi}\,{\mathscr D} \widetilde{\eta} \, \exp \Biggl[i \int_{t_0}^t d\tau d\chi d\bar{\chi} L\Bigl|_{\gamma=0}(\widetilde{\varphi},\widetilde{\eta})\Biggr].
\end{displaymath}

If we take into account also the term $\gamma \dot{\varphi}$ in (\ref{behind}) and we apply on it the dequantization procedure, then what we get is the derivative term $-\gamma \dot{\Lambda}_{\eta}=-\frac{d}{dt}(\gamma \Lambda_{\eta})$. This term does not play any dynamical role at the classical level, in the sense that it does not modify the classical equations of motion, so it can be disregarded, just like it has been disregarded in the implementation of the CPI by putting $\gamma=0$ from the beginning.

We can summarize this Letter by saying that here we have somehow obtained two further classical descriptions of spin. We have used the word ``somehow" because the descriptions we got are strictly related to the previously existing ones 
\cite{Casalbuoni}, \cite{Berezin}, \cite{nielsen}, \cite{vH}:
what we have built here is a sort of classical Lie derivative \cite{gozzi}, \cite{marsden} associated with the old descriptions of spin mentioned above. What instead is completely new in this Letter is the proof that the dequantization procedure proposed in \cite{deq} for non-spinning particles works also in the spinning case. 

\section*{Acknowledgments}
I would like to thank E. Gozzi for his precious help and suggestions.
This research has been supported by grants from INFN, MIUR and the University of Trieste.

\end{document}